\documentclass[12pt]{iopart}
\usepackage{amssymb}

\newcommand{\gsim}{\mbox{\raisebox{-.9ex}{~$\stackrel{\mbox{$$}}{\sim}$~}}}

\newcommand\vev[1]{{\langle {#1} \rangle}}

\newcommand\eq[1]{Eq.~(\ref{#1})}
\newcommand\eqs[2]{Eqs.~(\ref{#1}) and (\ref{#2})}

\newcommand\eqref[1]{(\ref{#1})}

\newcommand\pa{\partial}

\newcommand\ee{\end{equation}}
\newcommand\be{\begin{equation}}
\newcommand\eea{\end{eqnarray}}
\newcommand\bea{\begin{eqnarray}}

\newcommand\bfk{{\mathbf k}}

\newcommand\bfx{{\mathbf x}}

\newcommand\sub[1]{_{\rm #1}}

\newcommand{\fnl}{{f\sub{NL}}}

\begin{document}
\title{A bound concerning primordial non-Gaussianity}
\author{David H.\ Lyth and Ignacio Zaballa}\address{Physics department, Lancaster University,
Lancaster LA1 4YB, UK.} \ead{\mailto{d.lyth@lancaster.ac.uk},
\mailto{ i.zaballa@lancaster.ac.uk}}
\begin{abstract}
Seery and Lidsey have calculated the three-point correlator of the light
scalar fields, a few Hubble times after horizon exit during inflation.
Lyth and Rodriguez have calculated the contribution of this correlator to the
three-point correlator of the primordial curvature perturbation. We calculate
an upper bound on that contribution, showing that it is too small ever to be
observable.
\end{abstract}
\pacs{98.80.Cq \hfill JCAP10 (2005) 005, astro-ph/0507608 v2}

%
%
%
\section{Introduction}
The study of non-Gaussianity features in the primordial
curvature perturbation $\zeta$ has become a subject of growing interest,
because they provide a valuable discriminator between different models
for its origin \cite{brev,lr1,lr2,bl}.
While the relevant scales are outside the horizon, the curvature perturbation according to the $\delta N$ formalism
is given by \cite{ss,lms} (see also \cite{Starobinsky:1982ee,Hawking:1982cz,Guth:1982ec,star})
\bea
\zeta(\bfx,t) &=& \delta N(\phi_i(\bfx),\rho(t)) \label{deln1} \\
&=& \sum_i N_i(t) \delta\phi_i(\bfx) + \frac12
\sum_{ij} N_{ij}(t) \delta\phi_i(\bfx) \delta\phi_j(\bfx)
+ \cdots \label{deln2}
.
\eea
In this expression,  $N$ is the number of $e$-folds of expansion,
from a flat slice of spacetime on which the light fields during inflation
have values
$\phi_i(\bfx)=\phi_i+\delta\phi_i(\bfx)$,
and ending on a slice which has uniform energy density $\rho$.
The initial slice is taken to be a few $e$-folds after the relevant
scales have left the horizon. The final slice can be any time after $\zeta$
has settled down to the time-independent value which provides an initial
condition for the evolution of perturbations after horizon entry, and is
constrained by observation. We use the notation $N_i\equiv \pa N/
\pa \phi_i$ and $N_{ij}\equiv \pa^2 N/\pa \phi_i \pa \phi_j$, the derivatives
being evaluated with the fields at their unperturbed values $\phi_i$.

According to first-order cosmological perturbation theory, the field
perturbations
$\delta\phi_i(\bfx)$
are gaussian, with spectrum $(H/2\pi)^2$ where $H$ is the
Hubble parameter during inflation. (In this paper we ignore the scale
dependence of the spectrum.) Using this result, \eq{deln2} gives
the evolution of $\zeta$ without any further use of cosmological perturbation
theory.
The first term is Gaussian, and  higher terms are responsible for any
 non-gaussianity.

The question is whether it is is permissible to ignore the non-Gaussianity of
$\delta\phi_i$ which is generated at higher orders in cosmological perturbation theory. In this paper we answer the question in the affirmative, at least
for the next order in cosmological perturbation theory and for the
three-point correlator of $\zeta$. Our starting point is a recent
calculation of Seery and Lidsey \cite{sl}

%
\section{The three point correlator  and $f_{\rm NL}$}
Since observation shows that $\zeta$ is almost Gaussian, \eq{deln2}
must be dominated by the first term. The spectrum of the curvature
perturbation is therefore \cite{ss}
\begin{equation}\label{eq:curvaturepert}
\mathcal{P}_\zeta\simeq\left(\frac{H}{2\pi}\right)^2\sum^n_{i=1}N_{i}^2 \;,
\end{equation}

The three-point correlator of $\zeta$, or its bispectrum
 defined by $\vev{\zeta_{\bfk_1}\zeta_{\bfk_2}\zeta_{\bfk_3}}
=(2\pi)^3B_{\mathcal{\zeta}}\;\delta^3(\mathbf{k}_1+\mathbf{k}_2
+\mathbf{k}_3)$, is the lowest order signature of  non-Gaussianity.
Following Maldacena \cite{m}, we define
 $f_{\rm NL}(k_1,k_2,k_3)$  by
\begin{equation}\label{eq:bispectrum}
B_{\mathcal{\zeta}}\left(k_1,k_2,k_3\right)=-\frac{6}{5}f_{\rm
NL}\left[P_{\zeta}\left(k_1\right)P_{\zeta}\left(k_2\right)+\rm{cyclic}\right],
\end{equation}
where $P_{\zeta}(k)=2\pi^2\mathcal{P}_\zeta/k^3$.

At the level of first-order cosmological perturbation theory, the
the perturbations in the fields are Gaussian. Then  $f_{\rm NL}$
is almost scale-independent and given by  \cite{lr2,bl}
\begin{equation}\label{eq:fnlgaussian}\label{fnl}
-\frac{6}{5}f_{\rm NL}\simeq
\frac{\sum_{ij}N_{i}N_{j}N_{ij}}{\left(\sum_iN_{i}^2\right)^2}+\mathcal{P}_\zeta\frac{\sum_{ijk}N_{ij}N_{jk}N_{ki}}{\left(\sum_iN_{i}^2\right)^3}
\;.
\end{equation}

At the level of second-order cosmological perturbation,
the field perturbations  have non-Gaussianity which is specified entirely
by their three-point correlator. This adds to $\fnl$ the
following  contribution \cite{lr2}
\footnote{Other contributions will be studied in a separate paper, but are expected to be subdominant.}
\begin{equation}\label{eq:fnlnongaussian}
\Delta f_{\rm NL}=\frac{\sum_{i,j,k}N_{i}N_{j}N_{k}f^{ijk}_{\rm
NL}\left(k_1,k_2,k_3\right)}{\left(\sum_iN_{i}^2\right)^{3/2}\mathcal{P}_{\zeta}^{1/2}}
\;,
\end{equation}
where the
 $f^{ijk}_{\rm NL}\left(k_1,k_2,k_3\right)$
functions are related to the three point correlation functions of the
fields by
\bea \label{eq:bispectrumfields}
B^{ijk}_{\phi}\left(k_1,k_2,k_3\right)&=&-\left(4\pi^4\right)
\frac{6}{5}f^{ijk}_{\rm NL}\left(k_1,k_2,k_3\right)\;
\left(\frac{H}{2\pi}\right)^3\frac{\sum_ik_i^3}{\prod_ik_i^3} \;, \\
\label{eq:threepoint1}
\langle\delta\phi^i_{\mathbf{k}_1}\delta\phi^j_{\mathbf{k}_2}
\delta\phi^k_{\mathbf{k}_3}\rangle&=&
(2\pi)^3B^{ijk}_{\phi}\left(k_1,k_2,k_3\right)
\delta^3(\mathbf{k}_1+\mathbf{k}_2+\mathbf{k}_3). \eea

The quantity  $\fnl$ will eventually be observable  \cite{brev}
if $|\fnl|\gsim 1$. It will then be given accurately by \eq{fnl}, provided
that $|\Delta \fnl|\ll 1$.

Seery and Lidsey \cite{sl} find from
 second-order cosmological
perturbation theory
\begin{equation}\label{eq:threepoint2}
\vev{\delta\phi^i_{\mathbf{k}_1}\delta\phi^j_{\mathbf{k}_2}
\delta\phi^k_{\mathbf{k}_3}}=(2\pi)^3\frac{\delta^3(\sum_i\mathbf{k}_i)}
{\prod_ik_i^3}\frac{4\pi^4}{M_P^2}
\left(\frac{H}{2\pi}\right)^4\sum_{\sigma'}\frac{\dot\phi_i}{2H}\;
\delta_{jk}\mathcal{M}_{123} \;.
\end{equation}
Here the sum $\sigma'$ is over all the permutations of the indices i, j and k,
 at the same time their respective momenta $k_1$, $k_2$ and $k_3$, and
\begin{equation}\label{eq:Mfunction}
\mathcal{M}_{123}\left(k_1,k_2,k_3\right)=\frac{1}{2}\left(-3
\frac{k^2_2k_3^2}{k_t}-\frac{k^2_2k_3^2}{k_t^2}\left(k_1+2k_3\right)+
\frac{1}{2}k_1^3-k_1k_2^2\right),
\end{equation}
where $k_t=k_1+k_2+k_3$. Then  $\Delta f_{\rm NL}$  can be read
off from (\ref{eq:bispectrumfields}), (\ref{eq:threepoint1}) and
(\ref{eq:threepoint2}) resulting in
\begin{equation}\label{eq:fnlexpression}\label{delfnl}
-\frac{6}{5}\Delta f_{\rm
NL}=\frac{\sum_iN_{i}\left(\frac{\dot\phi_i}{2H}\right)\sum_{\sigma}\mathcal{M}_{123}}{M_P^2\sum_iN_{i}^2\sum_ik_i^3}
\;.
\end{equation}
The sum over $\sigma$ denotes the sum over the permutations of the three
momenta only.

If the only relevant field perturbation is that of the inflaton in a
single-component slow-roll model of inflation, the sum of \eqs{fnl}{delfnl}
reproduce \cite{sl} the result of Maldacena \cite{m}. In that case
$|\fnl|\ll 1$, making it too small to observe.

%
\section{The maximum of the $\Delta f_{\rm NL}$ function for constant $\zeta$.}
To maximise $\Delta f_{\rm NL}$ with fixed $\zeta$ we use the
Lagrange multipliers method. To do this, first write equation
(\ref{eq:fnlexpression}) in the following form
\begin{equation}\label{eq:fnlexpression2}
\frac{6}{5}\Delta f_{\rm
NL}\simeq\;\frac{\sum_iN_{i}V^i\;\mathcal{C}\left(k_1,k_2,k_3\right)}
{\sum_iN_{i}^2} \;,
\end{equation}
where we have used the slow roll condition $3H\dot\phi^i\simeq-V^i$, and $\mathcal{C}$\ is a function of the momenta only given by
\begin{equation}\label{eq:Cexpression}
\mathcal{C}\left(k_1,k_2,k_3\right)=\frac{1}{6H^2M_P^2}\frac{\sum_{\sigma}\mathcal{M}_{123}}{\sum_ik_i^3}
\;.
\end{equation}
Then, the differential of $\Delta f_{\rm NL}$ and the constraint
are respectively
\begin{eqnarray}
d\left(\Delta f_{\rm
NL}\right)=\frac{\mathcal{C}}{\sum_iN_i^2}\sum_iV^idN_i
\;,\label{eq:dfnl}\\
\sum_i2\lambda N^idN_i=0
\;,
\end{eqnarray}
where $\lambda$ is the Lagrange multiplier corresponding to the unique constraint of the problem, $\sum_iN_i^2=\rm{constant}$. Adding terms proportional to $dN_i$ we find that
\begin{equation}
N_i=-\frac{\mathcal{C}}{2\lambda}\frac{V_i}{\sum_iN_i^2}
\;.
\end{equation}
To find the value of $\lambda$ we need to add all the $N_i^2$'s. After doing this one gets
\begin{equation}
\lambda=\pm\frac{\mathcal{C}}{2}\frac{\left(\sum_iV_i^2\right)^{1/2}}{\left(\sum_iN_i^2\right)^{3/2}}
\;,
\end{equation}
and therefore the extrema of the function corresponds to the values
\begin{equation}
N_i=\pm\left(\sum_jN_j^2\right)^{1/2}\frac{V_i}{\left(\sum_jV_j^2\right)^{1/2}}
=\pm\frac{\mathcal{P}^{1/2}_{\zeta}}{\left(H/2\pi\right)}
\frac{V_i}{\left(\sum_jV_j^2\right)^{1/2}}
\;.
\end{equation}
To find the extrema of the non-linear function $f_{\rm NL}$ we
substitute the value above of $N_i$ in equation
(\ref{eq:fnlexpression2}), which yields
\begin{equation}\label{eq:fnlexpression3}
\frac{6}{5}|\Delta f_{\rm NL}|_{\rm max} =\frac{1}{6} \frac
{\vert\sum_{\sigma}\mathcal{M}_{123}\vert }{ \sum_i k_i^3} \frac
{\left(H/2\pi\right) }{ \mathcal{P}^{1/2}_{\zeta}M_P^2}
\sqrt{\sum_i\left(\frac{V_i}{H^2}\right)^2} \;.
\end{equation}
 Writing explicitly the terms in the sum
$\sum_{\sigma}\mathcal{M}_{123}/\sum_ik_i^3$ one gets
\begin{equation}\label{eq:Mfunction2}
\left| \frac{
\sum_{\sigma}\mathcal{M}_{123}\left(k_1,k_2,k_3\right)
}{
\sum_ik_i^3} \right|=\left|\frac{1}{2}-\frac{1}{2}
\frac{\sum_{i\neq j}k_ik_j^2}{\sum_ik_i^3}-4\frac{\sum_{i>j}
\frac{k_i^2k_j^2}{k_t}}{\sum_ik_i^3}\right|
\end{equation}
 The maximum value of the quantity in brackets is
 $11/6$, achieved for $k_1=k_2=k_3$.

The slow roll parameter $\epsilon$ is defined along the steepest descent
trajectory of the potential $V(\phi_i)$. That is,
\begin{equation}\label{eq:srollparameters}
\epsilon\equiv\frac{M_P^2}{2}\frac{\vert\vec{\nabla} V\vert^2}{V^2}=\frac{1}{18M_P^2}\sum_i\left(\frac{V_i}{H^2}\right)^2,
\end{equation}
Then, $\Delta f_{\rm NL}$ can be written in the following form:
\begin{equation}\label{eq:fnlexpression4}
\frac{6}{5}\vert\Delta f_{\rm NL}\vert_{\rm max}
=\frac{11}{12}\frac{\left(H/2\pi\right)}{\mathcal{P}^{1/2}_{\zeta}M_P}\sqrt{2\epsilon}.
\end{equation}
The power spectrum of gravitational waves,
$\mathcal{P}_{G}=8M_P^{-2}(H/2\pi)^2$, is independent of the number of
fields, and therefore we can express the tensor to scalar ratio $r$ as
\begin{equation}
r=\frac{\mathcal{P}_G}{\mathcal{P}_\zeta}=
\frac{8\left(H/2\pi\right)^2}{M_P^2\mathcal{P}_\zeta}.
\end{equation}
Introducing $r$ in equation (\ref{eq:fnlexpression4}) for $\Delta
f_{\rm NL}$, one gets
\begin{equation}\label{eq:fnlexpression5}
\frac{6}{5}\vert\Delta f_{\rm NL}\vert_{\rm max}
=\frac{11}{24}\sqrt{r\epsilon}.
\end{equation}

We can use a  recent analysis \cite{obs}
of observations to bound $r$ and $\epsilon$.
There is a direct bound  $r<0.46$. Also,  the bound
 $|n-1|<0.04$ on the spectral tilt, combined with the prediction
\cite{lrrev,book} $n-1= 2\epsilon + \cdots$ gives $\epsilon
\lesssim 0.02$ (barring an accurate cancellation in the last
formula). This gives
\begin{equation}\label{eq:fnlexpression6}
\frac{6}{5}|\Delta f_{\rm NL}| \lesssim 0.044.
\end{equation}
We conclude that the three-point correlator of $\zeta$ can safely be
calculated from \eq{eq:fnlgaussian}, if it is big enough to be observable.

\ack We thank the referee and Y. Rodriguez for pointing out
additional contributions to $\Delta f_{\rm NL}$. D.H.L. is
supported by PPARC grants PPA/G/O/2002/00469, PPA/V/S/2003/00104,
PPA/G/O/2002/00098 and PPA/S/2002/00272, and by EU grant
MRTN-CT-2004-503369. I.Z. is partially supported by Lancaster
University Physics Department.
%
%

\end{document}